\documentclass[pre,twocolumn,superscriptaddress, a4paper]{revtex4}

\usepackage{graphicx}
\usepackage{tabularx}
\usepackage{amsmath}
\usepackage{mathptmx}
\DeclareSymbolFont{myletters}{OML}{cmm}{m}{it}
\DeclareMathSymbol{\myeps}{\mathord}{myletters}{"0F}

\usepackage{hyperref}
\usepackage{titlesec}
\usepackage{xcolor}

\titlespacing{\section}{0pt}{12pt}{6pt}
\titlespacing{\subsection}{0pt}{12pt}{6pt} 

\newcolumntype{Y}{>{\centering\arraybackslash}X}

\begin{document}
\title{Temperature measurement from perturbations}
\author{Cheng Zhang}
\affiliation{Applied Physics Program and Department of Bioengineering, Rice University, Houston, TX 77005}

\begin{abstract}
The notion of configuration temperature is extended to discontinuous systems by identifying the temperature as the nontrivial root of several integral equations regarding the distribution of the energy change upon configuration perturbations.  The relations are generalized to pressure and a distribution mean force.
\end{abstract}

\maketitle

\newcommand{\eps}{\varepsilon}
\newcommand{\hk}{\mathrm{hk}}
\newcommand{\sgn}{\mathrm{sgn}}
\newcommand{\intn}{\mathrm{int}}
\newcommand{\vir}{\mathrm{vir}}
\newcommand{\la}{\langle}
\newcommand{\ra}{\rangle}
\newcommand{\LJ}{\mathrm{LJ}}
\newcommand{\hs}{\mathrm{hs}}
\newcommand{\vct}[1]{\mathbf{#1}}
\newcommand{\mat}[1]{\overset\leftrightarrow{#1}}

\section{Introduction}
While temperature is usually computed from the kinetic energy, 
%more recent studies showed 
it can also be measured from derivatives of the potential energy $U(\mathbf q)$  with respect to coordinates $\mathbf q$ \cite{rugh,butler,jepps}.  
In terms of the inverse temperature $\beta = 1/(k_B T)$, 
with $k_B$ being the Boltzmann constant, we have
\begin{equation}
\beta =
\left\langle 
\nabla \cdot \left( \frac{\nabla U}{\nabla U \cdot \nabla U} \right)
\right\rangle
\approx
\frac{\langle \nabla^2 U \rangle}
{\langle \nabla U \cdot \nabla U \rangle},
\label{eq:ctemp}
\end{equation}
where $\la\dots\ra$  denotes an ensemble average,
and $\beta$ 
%takes the respective meaning in various ensembles, 
%e.g., it 
is the logarithmic derivative of the density of states 
$\Omega(E)$ in the microcanonical ensemble:
$\beta_E = \frac{d}{dE} \log \Omega(E)$ 
\cite{mcquarrie, landau}, 
or the parameter in the canonical ensemble, etc.  
Except in the canonical ensemble 
\cite{jepps, mcquarrie, landau}, 
the last expression in Eq. (\ref{eq:ctemp}) has an $O(1/N)$ error in a system of $N$ degrees of freedom \cite{butler, jepps} (see also Appendix \ref{apd:ctemp}).
Eq. (\ref{eq:ctemp}) defines the so-called \emph{configuration temperature},
for it depends only on coordinates.
It has been used in verifying a Monte Carlo simulation \cite{butler}, constructing equilibrium \cite{yan} and non-equilibrium \cite{adib} microcanonical sampling, and building a thermostat \cite{braga}, etc.

As Eq. (\ref{eq:ctemp}) requires a continuous system with up to the second derivatives of the potential energy, 
it is inapplicable if the molecular potential is discontinuous \cite{butler, allen, frenkel}, 
or if the second derivatives are difficult to compute.  
Here we show that one can avoid the derivatives by computing the potential-energy change caused by a virtual perturbation, 
and obtain formulas of the configuration temperature that remain valid for irregular potential energy functions.

\section{Temperature from perturbations}

\subsection{\label{sec:mechtrans} Mechanical translation}

We begin with a mechanical translation of Eq. (\ref{eq:ctemp}) by a random perturbation.  
Given a configuration $\mathbf q$, 
a small perturbation $\mathbf u$ in coordinates changes the potential energy by 
$\eps = U(\mathbf{q+u}) - U(\vct{q}) 
\approx \nabla U \cdot \vct{u} + \frac{1}{2} \vct{u} \nabla \nabla U \cdot \vct{u}$.  
Now consider a random $\vct{u}$ that satisfies (a) the distribution of $\vct{u}$ is symmetric with equal probabilities for any $\vct{u}$ and $-\vct{u}$;  
(b) the components of $\vct{u}$ are independent: 
$\overline{u_i u_j} = \sigma^2 \delta_{ij}$, where the overline denotes an average over the perturbation $\vct{u}$, and $\sigma$ is the standard deviation of any $u_i$.  On average, we get 
$\overline{\eps}
\approx \frac{1}{2}\overline{
	\vct{u} \cdot \nabla \nabla U \cdot \vct{u} 
	}
= \frac 1 2 \nabla^2 U \sigma^2$
and 
$\overline{\eps^2} 
\approx \overline{(\nabla U \cdot \vct{u})^2}
= (\nabla U \cdot \nabla U) \sigma^2$  
to the leading order.  
Further averaging over $\vct{q}$ (denoted by $\langle \dots \rangle$) yields
\begin{equation}
\lim_{\sigma \rightarrow 0} \frac
{2 \langle \overline{\eps} \rangle}
{\langle \overline{\eps^2} \rangle }
=
\frac{\langle \nabla^2 U \rangle}
{\langle \nabla U \cdot \nabla U \rangle}
\approx
\beta.
\label{eq:mechtrans}
\end{equation}
%
%
%
%The last step follows from Eq. (\ref{eq:ctemp}).
%
%
The above perturbation need not to cover all coordinates, 
since Eq. (\ref{eq:mechtrans}) holds for a single coordinate: 
$\lim_{\sigma \rightarrow 0} 2 \langle \overline{\eps} \rangle \
/ \langle \overline{\eps^2} \rangle \
= \langle \partial_i^2 U \rangle/\langle (\partial_i U)^2 \rangle
\approx \beta $.  
The second step follows from a generalized formula 
$\beta \approx \langle \nabla \cdot \vct{B} \rangle/\langle \vct{B} \cdot U \rangle$ \cite{jepps} 
%(see also Appendix \ref{apd:ctemp}), 
%with the components of $\vct{B}$ given by 
with $B_i = \delta_{ij} U_j$.

Eq. (\ref{eq:mechtrans}) shows that the temperature can be obtained from the statistical moments, or, equivalently, the distribution, of the energy change $\eps$.
Note that a positive temperature $\beta$ demands an asymmetry $\langle \overline{\eps}\rangle > 0$
from the symmetric perturbation.  
This reflects the fact that the entropy increases with the energy,
which induces, on average, 
a convex potential-energy surface \cite{weinhold} that favors a positive $\eps$.  %[Fig. \ref{fig:scheme}(a)].

As Eq. (\ref{eq:mechtrans}) requires no derivative of the potential energy function $U(\vct{q})$,
we expect it to be applicable to a discontinuous $U(\vct{q})$.  
Nonetheless, the perturbation still needs to be rather small, $\eps \ll O(\ref{eq:ctemp})$.
Below we derive more practical integral relations for an $\eps$ as large as $O(\ref{eq:ctemp})$.

\subsection{Canonical ensemble}

Consider a canonical ensemble.  
If each configuration $\vct{q}$ is shifted by $\vct{u}$ to $\vct{q}' = \vct{q+u}$, 
then the potential-energy change $\eps = U(\vct{q}') - U(\vct{q})$ satisfies \cite{maes, voter, jarzynski2002}
\begin{align}
  \langle \exp(-\beta \eps) \rangle_\beta 
  &=  \int \exp(-\beta \eps) \, 
  e^{ -\beta U(\vct{q})}/Z(\beta) \, d\vct{q} \notag\\
  &=  \int e^{ -\beta U(\vct{q}')} / Z(\beta) \, d\vct{q}' = 1,
\label{eq:canon}
\end{align}
where $Z(\beta) \equiv \int e^{-\beta U(\vct{q})} d\vct{q}$ is the partition function. 
%and $e^{-\beta U(\vct{q})}/Z(\beta)$ is the ensemble weight for configuration $\vct{q}$.  
Thus, the equation $\la \exp(-\beta' \eps) \ra_\beta =1$ 
%of the variable $\beta'$ 
has a nontrivial root at $\beta' = \beta$.

Eq. (\ref{eq:canon}) yields the same $\beta$ value as 
Eq. (\ref{eq:mechtrans}) in the limit of small perturbation.  
%If $\eps$ is small, 
First,
since
$\log \langle \exp(-\beta \eps)\rangle_\beta 
\approx
-\beta \langle \eps \rangle_\beta
+ \frac 1 2 \beta^2 \langle \Delta \eps^2 \rangle_\beta$,
where $\Delta \eps = \eps -\langle \eps \rangle_\beta$, we have
%The only solution, apart from $\beta = 0$, is
\begin{equation}
\beta \approx 2 \la \eps \ra_\beta / \la \Delta \eps^2 \ra_\beta,
\tag{$\ref{eq:canon}'$}
\label{eq:gaussian}
\end{equation}
which is reduced to Eq. (\ref{eq:mechtrans}) after averaging over perturbations:
$\overline {\la \eps \ra}_\beta$
is $O(\sigma^2)$, 
while $\eps$ is $O(\sigma)$, 
so $\eps \approx \Delta \eps$.
As Eq. (\ref{eq:gaussian}) is also exact 
for a Gaussian $\eps$ distribution, 
a perturbation involving many similar degrees of freedom 
can assume a larger $\eps$.

We can replace the above continuous configuration $\vct{q}$
	by a discrete one \cite{maes}, 
and/or the perturbation $+\vct{u}$ 
by an invertible transformation 
	from $\vct{q}$ to $\vct{q}'$ with a unit Jacobian:
$|\partial \vct{q}'/\partial \vct{q}| = 1$ \cite{jarzynski2002}.  
Some generalizations of Eq. (\ref{eq:canon}) are discussed 
	in Appendix \ref{apd:fluc}, and verified on a harmonic oscillator (Appendix \ref{apd:ho}).
The requirement for a unit Jacobian should, however, be relaxed in the non-equilibrium case (Appendix \ref{apd:langevin}).

%\begin{figure}[b]
%  \begin{minipage}{\linewidth}
%    \begin{center}
%        \includegraphics[angle=0, width=\linewidth]{../fig/pt_scheme_color.png}
%    \end{center}
%  \end{minipage}%
%  \caption{
%(a) The intersection of the sphere (dotted) of perturbations
% $\vct{u}$ (isotropic, fixed length)
%with the constant potential-energy surface (dashed) at $U+\eps$
%has a larger circumference than that with the surface at $U-\eps$.  
%Thus, $+\eps$ occurs more often than $-\eps$, even for a symmetric $\vct{u}$.  
%(b) The forward and reverse fluxes (arrows), 
%e.g., $\Phi(U, U+\eps)$ and $\Phi(U+\eps, U)$, 
%are balanced between any two $U$ ensembles (ovals).  
%The flux $\Phi(U, U+\eps)$ is proportional to the ensemble population $g(U)$
%with the ratio $p_U(\eps)$ changing slowly with $U$.
%(c) The distribution $p_U(\eps)$ (solid) of the energy change $\eps$ upon perturbations.  
%With the correct $\beta$,  
%$\exp(-\beta \eps) \, p_U(\eps)$ (dashed) roughly coincides with the horizontal reflection $p_U(-\eps)$, 
%and thus satisfies $\la \exp(-\beta\eps) \ra_U = 1$; it is not so with another $\beta'$ (dot-dashed).
%}
%\label{fig:scheme}
%\end{figure}

\subsection{\label{sec:uensemble}Constant potential-energy ensemble}

We now adapt Eq. (\ref{eq:canon}) to a constant potential-energy ensemble, or a $U$ ensemble below, 
which collects configurations with the same potential energy $U$.  
The ensemble is commonly used to build a multicanonical ensemble of a flat potential-energy distribution \cite{yan, multicanonical}.  
Configurations of the $U$ ensemble sum to the density of potential-energy states $g(U) = \int \delta[U(\vct{q} - U]\, d\vct{q}$, 
and the temperature of potential energy is defined as $\beta_U(U) = \frac{d}{dU} \log g(U)$.  
We will show
\begin{equation}
\la (-\eps)^k e^{-\beta_U(U)\, \eps} \ra_U
\approx \la \eps^k \ra_U,
\label{eq:expu}
\end{equation}
for a nonnegative integer $k$,
where $\la\dots\ra_U$ denotes an average over the surface $U = U(\vct{q})$. 
Particularly, $\la \exp(-\beta_U \, \eps) \ra_U \approx 1$.

If the perturbation $\vct{u}$ is symmetric
(i.e., $+\vct{u}$ and $-\vct{u}$ are equally likely),  
then for each $\vct{u}$ that carries $\vct{q}$ to $\vct{q}' = \vct{q+u}$, 
the inverse $-\vct{u}$ that carries $\vct{q}'$ back to $\vct{q}$ shares the same probability.  
Without any \textit{a priori} bias of the start-point configuration,
the overall flux $\Phi(U, U+\eps)$, 
or the total number of $\vct{u}$ that reach the $U+\eps$ ensemble from the $U$ ensemble, 
therefore, equals the reverse flux $\Phi(U+\eps, U)$, 
or $\Phi(U, U+\eps) = \Phi(U+\eps, U)$.
%[Fig. \ref{fig:scheme}(b)].
%
But $\Phi(U, U+\eps)$ is the product of the density of states $g(U)$ 
and the $\eps$ distribution $p_U(\eps)$ for perturbations that start from the $U$ ensemble; so
\begin{equation}
g(U) \, p_U(\eps) = g(U+\eps) \, p_{U + \eps} (-\eps).
\label{eq:dbu}
\end{equation}
If $p_U(\eps)$ changes slowly with $U$: $p_{U+\eps}(-\eps) = p_U(-\eps)$, then
\begin{equation}
\frac{p_U(-\eps)}{p_U(+\eps)}
\approx
\frac{ g(U) }{ g(U+\eps) }
\approx
e^{-\beta_U (U) \, \eps},
\label{eq:probrev}
\end{equation}
and
\begin{align}
\la e^{(\beta' - \beta_U) \eps} \ra_U
&= \int_{-\infty}^\infty e^{(\beta' - \beta_U) \eps} \, p_U(\eps) d \eps \notag \\
&\approx \int_{-\infty}^\infty e^{(-\beta')(-\eps)} \, p_U(-\eps) d \eps 
= \la e^{ -\beta' \eps} \ra_U, 
\label{eq:expubp}
\end{align}
for any  $\beta'$.
By taking derivatives with respect to $\beta'$, 
and then setting $\beta' = 0$, we get Eq. (\ref{eq:expu}).  
%Eqs. (\ref{eq:probrev}) and (\ref{eq:expubp}) also have the exact counterparts in the canonical ensemble (Appendix \ref{apd:fluc}).

%Geometrically, Eq. (\ref{eq:probrev}) shows that the regions A and B
%in Fig. \ref{fig:scheme}(c) are roughly congruent.  
%Thus, we also have
Alternatively,
we can integrate Eq. (\ref{eq:probrev})
with the Metropolis acceptance probability as
\begin{equation}
\la \sgn(\eps) |\eps|^k	 \min\{1, e^{-\beta_U(U) \, \eps} \} \ra_U \approx 0,
\label{eq:symu}
\end{equation}
for a nonnegative $k$.
The $k = 1$ version shows that the energy change 
after a Metropolis step from the $U$ ensemble averages to zero 
if the parameter $\beta$ in the acceptance probability $\min\{1, \exp(-\beta \, \eps)\}$ roughly matches $\beta_U(U)$.  
Besides, the function $\exp(-\beta'\, \eps) p_U(\eps)$ is roughly even at $\beta' = \frac{1}{2} \beta_U(U)$.  Thus, for a nonnegative $k$
\begin{equation}
\left\la 
	\sgn(\eps) |\eps|^k 
	e^{ - \beta_U(U) \, \eps/2 } 
\right\ra_U \approx 0.
\label{eq:halfu}
\end{equation}

Eq. (\ref{eq:expu}) has an $O(1/N)$ error,
and can be corrected as
\begin{equation}
\beta_U \approx \hat \beta 
+ \dfrac{1}{2} \dfrac{\la \eps^{k+2} \ra_U} { \la \eps^{k+1} \ra_U } \dfrac{d \hat \beta} {d U}
- \dfrac{d \log \la \eps^{k+1} \ra_U} {d U},
\label{eq:expuc}
\end{equation}
where $\hat \beta$ is the $\beta_U$ value determined from Eq. (\ref{eq:expu}) (as an equality). 
%$\la (-\eps)^k \exp(-\hat \beta \, \eps)\ra_U = \la \eps^k \ra_U$. 
The correction reduces the error of $\beta_U$ to $O(1/N^2)$ if $\eps$ is $O(\ref{eq:ctemp})$,
such that the error of $\log g(U) = \int^U \beta_U(U') \, d U'$, 
integrated over an $O(N)$ domain, is $O(1/N)$.
Eq. (\ref{eq:expuc}) is also exact for any $N$ in the limit of small $\eps$ (Appendix \ref{apd:series}). 
The derivatives with respect to $U$ are readily computed
% from data collected at a neighboring $U$ 
in a usual simulation that allows the potential energy to fluctuate.  
The corrections for Eqs. (\ref{eq:symu}) and (\ref{eq:halfu}) can be found in Appendix \ref{apd:series}.

\subsection{\label{microcanonial}Microcanonical ensemble}

The temperature $\beta_E$ in the microcanonical ensemble can be obtained by substituting the total energy $E$ for the potential energy $U$ in the formulas for the $U$ ensemble; e.g., Eq. (\ref{eq:expu}) becomes 
$\la (-\eps)^k \exp(-\beta_E \eps ) \ra_E \approx \la {\eps}^k\ra_E$.  
The perturbations can still be limited to a subset of degrees of freedom, which can be made of only coordinates, but not momenta.
The limit $E \rightarrow \infty$, however, eliminates the $O(1/N)$ corrections in, e.g., Eq. (\ref{eq:expuc}).
Thus, Eqs. (\ref{eq:expu}), (\ref{eq:probrev})-(\ref{eq:halfu}) are exact if the interaction is weak compared to the kinetic energy, or if the system is embedded in a much larger isolated reservoir, which constitutes a canonical ensemble (see also Appendix \ref{apd:fluc}).

%The use of the exponential function generally renders
%Eqs. (\ref{eq:expu}), (\ref{eq:symu}) and (\ref{eq:halfu}) 
%exact only in the canonical ensemble
%(with $\la\dots\ra_U \rightarrow \la \dots \ra_\beta$).
%The use of the exponential function are intrinsically associated with the canonical ensemble. 
We may therefore interpret the relations as virtual ``thermometers'' 
gauged in the canonical ensemble,  
e.g., the average $\la \exp(-\beta \eps) \ra$  
invariably reads 1.0 in the canonical ensemble of the correct $\beta$
for any uniform perturbation, 
but not so in other ensembles.
It is, however, possible to construct an exact thermometer gauged in the microcanonical ensemble.  
The momenta-averaged weight for a configuration, whose potential energy is $U$, is $w(U) \propto (E-U)^{N/2 - 1}/\Omega(E)$ for a system of  $N$ degrees of freedom \cite{yan, martin-mayor}. 
%where $\Omega(E)$ is the density of state.  
Thus, the total energy $E$ can be estimated from the root of  
$\la [1 - \eps/(E-U)]^{N/2-1} \ra \approx 1$;
then
% the temperature is given by 
$\beta_E 
= \frac{d}{dE}\log \Omega(E) 
= \la (N/2 -1)/(E-U)\ra$ \cite{rugh}.

\section{Generalizations}

\subsection{Pressure}

The extension to pressure $p$ is straightforward.  
For a virtual volume move 
\cite{eppenga}
induced by the scaling of coordinates 
$\vct q \rightarrow (V'/V)^{1/D} \vct q$ 
(the dimension $D = 3$), we have \cite{jarzynski}
\begin{equation}
\la \exp[\Delta S -\beta (\Delta U + p \Delta V)]
\ra = 1,
\label{eq:pressure}
\end{equation}
where $\Delta U = U_{V'}(\sqrt[D]{V'/V}\vct{q}) - U(\vct{q})$,
and $\Delta V = V' - V$.  
The $\Delta S$ is computed from either the Jacobian of the coordinate scaling as
$\Delta S = (N/D) \log(V'/V)$,
or from the kinetic energy change as
$\Delta S = -[(V'/V)^{2/D} - 1]
\, \beta \, \vct{p}^2/(2m)$
of the conjugate momentum scaling 
$\vct{p} \rightarrow (V/V')^{1/D} \vct{p}$
\cite{andersen, hoover}.  
Here, $N$ is the number of degrees of freedom, 
not the number of particles.
For an infinitesimal volume change,
	we have $p = \la p_{\intn} \ra$, where 
$p_{\intn} = [(N/\beta) + \vir]/(DV)$ 
or
$p_{\intn} = (\vct{p}^2/m + \vir)/(DV)$
with 
$\vir = -\vct{q} \cdot \nabla U 
- (DV)(\partial U/\partial V)$.

Eq. (\ref{eq:pressure}) holds for a fixed \cite{eppenga} or variable $\Delta V$.  For a hard-core system, only negative $\Delta V$ should be used \cite{frenkel} 
%for the low-density configuration space is strictly larger than the high-density one  
(similar to Widom's method \cite{widom} for estimating the chemical potential, in which particles are inserted, but not removed).  
If $V$ is changed on an exponential scale as  
$V' = V e^\delta$ 
%(which avoids a negative $V'$)
\cite{frenkel},
then an additional factor $V'/V$ should be inserted into the brackets $\la \dots \ra$.
The advantage of using a variable volume change is that the virtual volume move is identical to 
a generalized (or possibly optimized \cite{jarzynski2002})
MC volume trial in an isothermal-isobaric \cite{frenkel} or Gibbs-ensemble \cite{gibbsensemble} MC simulation, and thus can be realized simultaneously.  
The formulas are exact in the isothermal-isobaric ensemble, but approximate in general.

\subsection{One-dimensional distribution mean force}

The formulas for temperature can be generalized to compute the logarithmic derivative, or the mean force, 
of a distribution $\rho(X)$  of an extensive quantity 
$X = X(\vct{q})$ \cite{rugh, meanforce, bluemoon, darve2008, zhang}:
\begin{equation}
\rho(X)
= \int \delta[X - X(\vct{q})] \, w(\vct{q}) \, d\vct{q},
\label{eq:distr1d}
\end{equation}
where $w(\vct{q})$ is the ensemble weight for a configuration $\vct q$, e.g., 
$w(\vct{q}) \propto \exp[-\beta U(\vct{q})]$
in the canonical ensemble.  
We wish to find $f(X) = \frac{d}{dX} \log \rho(X)$,
whose integral yields the free energy \cite{meanforce, bluemoon, darve2008}
$\beta F(X) = -\log \rho(X)$.

We define an \emph{adjusted} perturbation $\vct{u}^*$ as a symmetric perturbation $\vct u$
adjusted by the Metropolis acceptance probability:
\begin{equation}
A(\vct{q} \rightarrow \vct{q} + \vct{u})
= 
\min\{
1,
w(\vct{q} + \vct{u})/w(\vct{q})
\}.
\label{eq:adjpert}
\end{equation}
Thus, $\vct{u}^*$
is equal to $\vct u$ 
if the unadjusted perturbation $\vct u$  is accepted, 
or \textbf{0} if rejected.  
Like an MC move \cite{allen, frenkel, newman}, 
	the $\mathbf{u}^*$ satisfies detailed balance:
$w(\vct q) \, \pi(\vct{q} \rightarrow \vct{q}')
=
w(\vct{q}') \, \pi(\vct{q}' \rightarrow \vct{q})$,
where $\vct{q}' = \vct{q} + \vct{u}^*$. 
%and $\pi(\vct{q} \rightarrow \vct{q}')$ is the adjusted transition probability.  
Summing over $\vct{q}$ and $\vct{q}'$ 
under $X(\vct{q}) = X$  and 
$X(\vct{q}') = X + \eps$ leads to
\begin{equation}
\rho(X) \, p_X(\eps) = \rho(X+ \eps) \, p_{X+\eps}(-\eps).
\label{eq:dbx}
\end{equation}
Here, $\rho(X)$ replaces $g(U)$ in Eq. (\ref{eq:dbu}).
The temperature $\beta_U = \frac{d }{dU} \log g$ is then mapped to the mean force  $f = \frac{d }{dX} \log \rho$, and
\begin{equation}
\la \exp[-f(X) \, \eps] \ra_X \approx 1,
\label{eq:expx}
\end{equation}
where 
$\la \dots \ra_X$ denotes the $w(\vct q)$ weighted configuration average at a fixed $X$.
Note that $\eps$  is the change of $X$  after the adjustment, which is 0 for a rejected perturbation.
Similarly, the counterparts of Eqs. (\ref{eq:symu}) and (\ref{eq:halfu}) [and the $1/{N}$ corrections
(\ref{eq:expuc}), (\ref{eq:symuc}) and (\ref{eq:halfuc})] can be obtained by 
$U \rightarrow X$ and 
$\beta_U \rightarrow f$.

The above MC-type adjusted perturbation can be replaced by a reversible time evolution \cite{maes}, 
as long as the latter also satisfies Eq. (\ref{eq:dbx}).
We can therefore treat a short segment of such a trajectory as an adjusted perturbation, 
and use Eq. (\ref{eq:expx}) with
$\eps = X(t+\tau) - X(t)$ 
($\tau$ is the segment length).
Further, since Eq. (\ref{eq:expx}) does not require the potential energy, 
it can be readily applied to trajectories of 
colloidal particles monitored by microscopy experiments \cite{experiments}.

\subsection{Multidimensional distribution mean force}

We now consider a multidimensional distribution
\begin{equation}
\rho(\vec{X})
= 
\int \textstyle\prod_\alpha 
\delta[X_\alpha - X_\alpha(\vct{q})]
\, w(\vct{q}) \, d\vct{q}
\label{eq:distrmd}
\end{equation}
of $K$ extensive quantities
$\vec{X} = \{ X_\alpha \}$,
$\alpha = 1, \dots, K$.  
To find all
$\partial \log \rho/{\partial X_\alpha}$, 
we can extend, e.g., Eq. (\ref{eq:expu})
with $k = 1$, as
\begin{equation}
\left\la 
\eps_\alpha 
\exp \left( - \textstyle\sum_\gamma \hat f_\gamma \eps_\gamma \right) + \eps_\alpha 
\right\ra_{\vec{X}} = 0,
\label{eq:expxmd}
\end{equation}
where $\eps_\alpha = \Delta X_\alpha$ is the change of $X_\alpha$ by the adjusted perturbation, 
and $\la \dots \ra_{\vec X}$
denotes an average in the ensemble at a fixed set of  $\vec{X} = \{X_\mu\}$.  
Since Eq. (\ref{eq:expxmd}) offers $K$  equations, all components 
$\hat f_\gamma$  can be determined.  Thus,  
${\partial \log \rho}/{\partial X_\alpha}  \approx \hat f_\alpha$ 
to the first order, and the $O(1/N)$  correction [cf. Eq. (\ref{eq:expuc})] is
\begin{align}
\frac{\partial \log \rho}{\partial X_\alpha}
&\approx
\hat f_\alpha
-
\textstyle\sum_{\gamma \theta}
\big(\mat{M}^{-1}\big)_{\alpha \gamma}
\partial \la \eps_\gamma \eps_\theta \ra_{\vec X}
/
\partial{X_\theta}
\notag\\
&+\frac{1}{2} \textstyle\sum_{\gamma \theta \mu}
\big(\mat{M}^{-1}\big)_{\alpha \gamma}
(\partial \hat f_\mu/\partial X_\theta)
\la \eps_\gamma \eps_\mu \eps_\theta \ra_{\vec X},
\label{eq:expxmdc}
\end{align}
where $\mat{M}^{-1}$ is the inverse of 
the matrix 
$\big(\mat{M}\big)_{\alpha \gamma} = \la \eps_\alpha \eps_\gamma \ra_{\vec X}$.

The small-perturbation limit of
Eqs. (\ref{eq:expxmd}) and (\ref{eq:expxmdc})
is free from the $O(1/N)$ error,
and identical to the exact relation given by 
Eqs. (\ref{eq:mfmd}) and (\ref{eq:mfmd2}).
First, according to the definition of the adjusted perturbation Eq. (\ref{eq:adjpert}), 
we have $\eps_\alpha \approx 
(\nabla X_\alpha \cdot \vct{u}
+ \frac{1}{2} \vct{u} \cdot 
\nabla \nabla X_\alpha \cdot \vct{u})
\,
\min\{1, \exp(\nabla \log w\cdot \vct{u})\}$
(cf. Sec. \ref{sec:mechtrans}).  
Thus,
$\la \eps_\alpha \ra_{\vec X}
\approx \frac{1}{2} \sigma^2
\la \nabla^2 X_\alpha 
+ \nabla X_\alpha \cdot
	\nabla \log w
\ra_{\vec X}
=
\frac{1}{2} \sigma^2 F_\alpha$
(the acceptance probability only affects half of the perturbations that point against the gradient of $w$)
and
$\la \eps_\alpha \eps_\gamma \ra_{\vec X}
\approx \sigma^2 
\la 
\nabla X_\alpha \cdot \nabla X_\gamma 
\ra_{\vec X} = \sigma^2 W_{\alpha \gamma}$
to the leading order 
[we have averaged over symmetric perturbations and used Eq. (\ref{eq:mfmd2})].
For small $\eps_\alpha$, 
Eq. (\ref{eq:expxmd}) becomes 
$\sum_\gamma \la \eps_\alpha \eps_\gamma\ra_{\vec X}
\hat f_\gamma \approx 
2\la \eps_\alpha \ra_{\vec X}$, 
or 
$\sum_\gamma \sigma^2 W_{\alpha \gamma} \hat f_\gamma
\approx \sigma^2 F_\alpha$.  
Eq. (\ref{eq:expxmdc}) is then reduced to
$
\frac{\partial}{\partial X_\alpha}
\log \rho
\approx
\hat f_\alpha
-
\sum_{\gamma \theta} 
(\mat W^{-1})_{\alpha \gamma}
\partial W_{\gamma \theta}/\partial X_\theta$,
which is Eq. (\ref{eq:mfmd}) 
[the
$\la \eps_\gamma \eps_\mu \eps_\theta\ra_{\vec{X}}$
term is $O(\sigma^4)$, hence negligible].

\begin{figure*}[t]
  \begin{minipage}{0.75 \linewidth}
    \begin{center}
        \includegraphics[angle=-90, width=\linewidth]{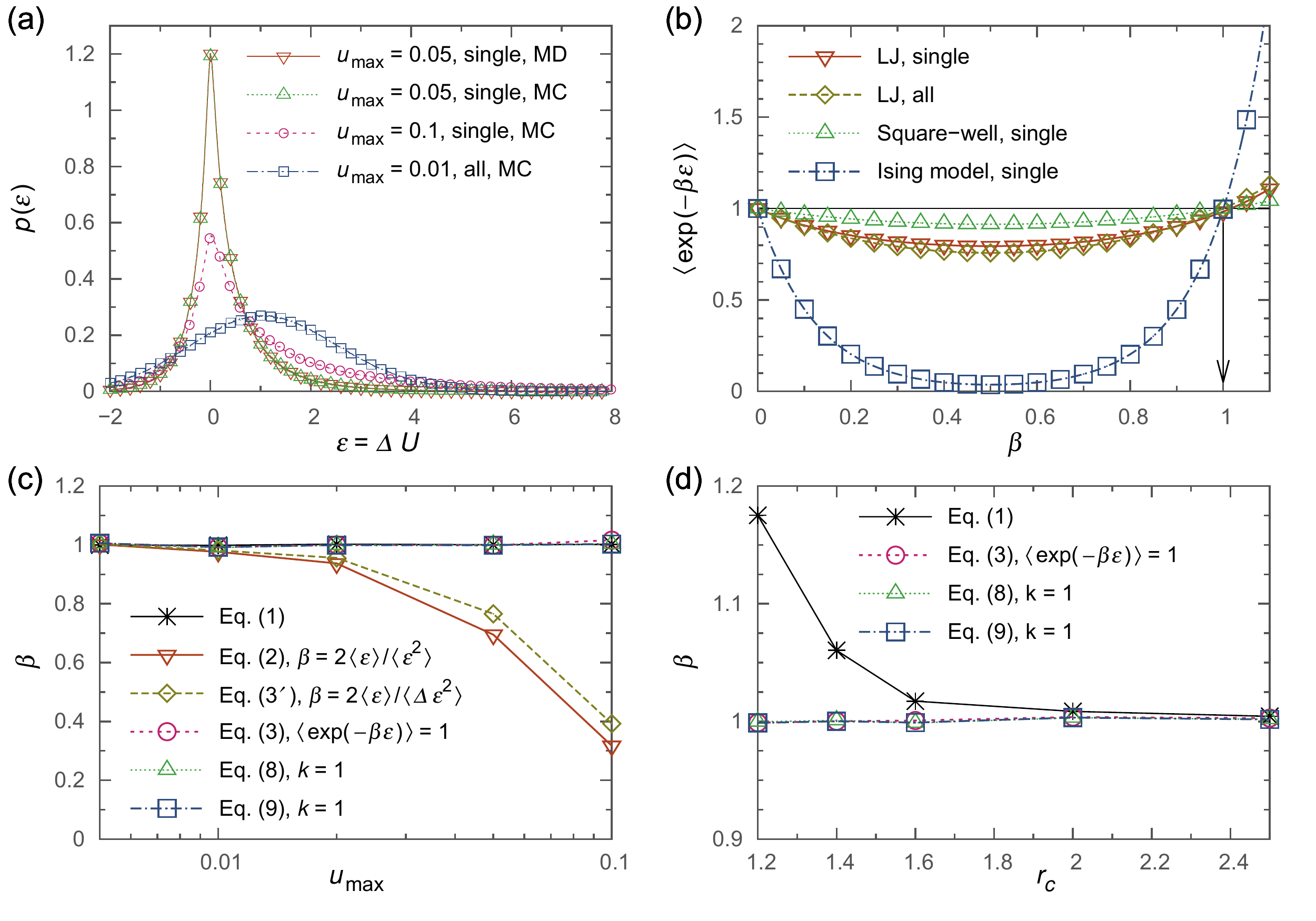}
    \end{center}
  \end{minipage}
  \caption{(a) The distributions of the potential-energy change $\eps$.
  (b) The intersection of $\exp(-\beta \eps)$ and 1.0 gives the correct $\beta$.
  (c) The temperature versus the perturbation amplitude $u_{\max}$.
  (d) The temperature versus the potential cutoff $r_c$.
  }
  \label{fig:basic}
\end{figure*}

\section{Numerical results}

We show some results for the temperature formulas.  
We use Eqs. (\ref{eq:expu}), 
(\ref{eq:symu}) or (\ref{eq:halfu}), 
etc., in an MC/MD  simulation in this way: 
once every few steps along the trajectory, 
we perturb the current configuration $\vct{q}$ by a random $\vct{u}$ 
(which is conducted as a virtual displacement 
so as not to disturb the real trajectory), 
and register the resulting change 
$\eps = U(\vct{q} + \vct{u}) - U(\vct{q})$
in the potential energy; 
the formulas then estimate the temperature $\beta$ 
from the accumulated distribution $p(\eps)$.
%Since no derivative of $U(\vct{q})$ is needed, 
%the formulas can be applied to systems of discontinuous potentials.   
%Further, we allow discrete configurations $\vct{q}$ and perturbations $\vct{u}$, 
%e.g., in 
For the Ising model, 
$\vct{q}$ is a spin configuration, 
and $+\vct{u}$ means to flip a random spin.

\subsection{Canonical and microcanonical ensembles}

Four $\eps$-distributions $p(\eps)$ from simulations on the 108-particle Lennard-Jones (LJ) fluid at $\rho = 0.7$ and $\beta_0 = 1.0$ are shown in Fig. \ref{fig:basic}(a).  
The first simulation used a regular MD (in the microcanonical-like ensemble), 
while the others used the Metropolis MC (in the canonical ensemble).  
The pair potential was switched smoothly from the standard form
$u(r) = 4 \myeps_{LJ} [(\sigma_{\LJ}/r)^{12} 
- (\sigma_{\LJ}/r)^6]$
($\myeps_{\LJ} = \sigma_{\LJ} = 1$)
at $r_s = 2.0$ to a 7th order polynomial 
%at which point the polynomial and first %three derivatives vanished 
\cite{zhang}
that vanishes at $r_c = 2.5$
to avoid artifacts.  
In the first three cases, each coordinate of a random particle was displaced by a random number in $(-u_{\max}, u_{\max})$, 
with $u_{\max} = 0.05$ (the first two cases) or 0.1 (the third).  
In the last case, the perturbation was applied to all particles
with $u_{\max} = 0.01$.  
The number of perturbations was $10^7$ in each case.  
Eqs. (\ref{eq:symu}) and (\ref{eq:halfu}) were adapted to the MD or canonical ensemble
in the sense of $\la \dots \ra_U \rightarrow \la \dots \ra$.
A comparison of the first two cases shows that the distribution was not very sensitive to the ensemble type in this case.
%whether the ensemble was microcanonical-like (MD) or canonical (MC).  
The distributions from the single-particle perturbations peaked at $\eps = 0$ [cf. Eq. (\ref{eq:hope})].  
The distribution from the all-particle perturbation, however, was Gaussian-like [cf. Eq. (\ref{eq:hogaussian})].  
The integral relations applied to all cases, e.g., the $\beta$ given by Eq. (\ref{eq:canon}) were 1.0077, 0.9990, 1.0160, and 0.9997, respectively.

The solution process of Eq. (\ref{eq:canon}) is illustrated in Fig. \ref{fig:basic}(b): 
if we plot $\la \exp(-\beta \eps)\ra$  against $\beta$, 
its intersection with 1.0 then gives the desired  $\beta$.  Three systems, the LJ fluid (with both single- and all-particle perturbations), 
square-well \cite{allen} fluid and 
$32\times 32$ Ising model were simulated 
in the canonical ensemble at $\beta_0 = 1.0$ 
using the Metropolis algorithm.  
%The number of perturbations was $10^7$ in each case, 
%and the computed $\beta$ were 
%1.0160 (LJ, single-particle), 
%0.9997 (LJ, all-particle), 
%1.0003 (square-well), 
%and 1.0004 (Ising), respectively.  
The square-well potential of two particles is infinity if their distance $r < r_a$, 
or $-\myeps_{\mathrm{sq}}$ if $r_a \le r < r_b$, 
or 0 otherwise 
($r_a =1$, $r_b = 1.5$, 
and $\myeps_{\mathrm{sq}} = 1$).  
%The single-particle perturbation with $u_{\max} = 0.1$ was used.  
%The density $\rho = 0.7$.  
Perturbations that produced clashes, hence infinite $\eps$, were excluded
from entering Eq. (\ref{eq:canon}). 
%for clashed configurations do not contribute to the ensemble.

In Fig. \ref{fig:basic}(c), the temperatures computed from 
Eqs. (\ref{eq:ctemp})-(\ref{eq:gaussian}), 
(\ref{eq:symu}) and (\ref{eq:halfu}),
%($k = 1$ in the last two cases), 
were plotted against $u_{\max}$, 
the size of the single-particle perturbation.  
All formulas worked well with a small $u_{\max}$, 
but with a large $u_{\max}$, 
Eq. (\ref{eq:mechtrans}) and (\ref{eq:gaussian})
quickly lost accuracy, 
while the integral formulas were little affected.
%[Eq. (\ref{eq:gaussian}) would, however, perform better for the all-particle perturbation, 
%whose distribution $p(\eps)$ was Gaussian-like].

The effect of the potential truncation \cite{butler} was studied using a set of MC simulations 
($u_{\max} = 0.05$) 
with the LJ potential deliberately truncated at small $r_c$.  
The discontinuity rendered Eq. (\ref{eq:ctemp}) approximate \cite{butler}.  
As shown in Fig. \ref{fig:basic}(d), 
a small $r_c$ indeed affected Eq. (\ref{eq:ctemp}),
but not Eqs. (\ref{eq:canon}), (\ref{eq:symu}), 
and (\ref{eq:halfu}).

\subsection{Constant potential-energy ensemble}

\begin{figure*}[t]
  \begin{minipage}{0.7 \linewidth}
    \begin{center}
        \includegraphics[angle=-90, width=\linewidth]{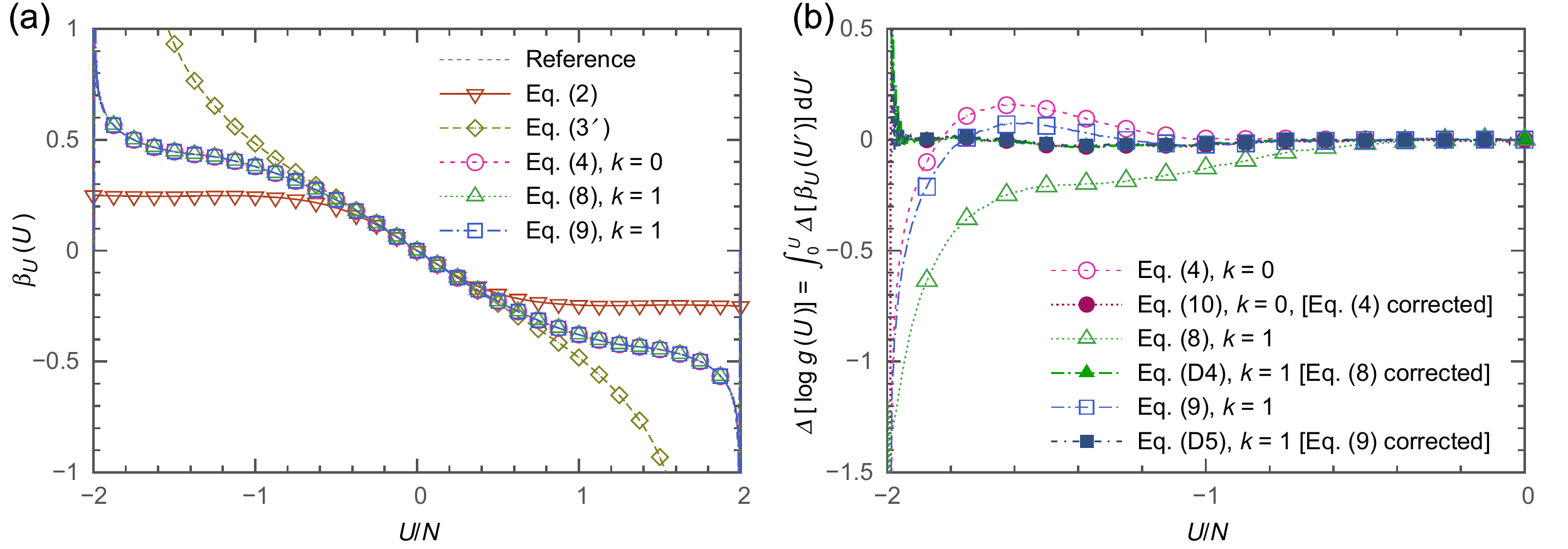}
    \end{center}
  \end{minipage}%
  \caption{(a) The profile of the temperature of potential energy  $\beta_U(U)$ along $U$  in the $32\times 32$ Ising model.  
(b) The error of $\log g(U) = \int_0^U \beta_U(U') \, dU'$.  
The reference values $\beta_U^*(U)$ were computed as 
$[\log g^*(U+\Delta U) - \log g^*(U - \Delta U)]/(2 \Delta U)$ 
with $\Delta U = 4$.
}
\label{fig:isent}
\end{figure*}

The temperature profile $\beta_U(U)$ along the potential energy was computed on the $32\times 32$ Ising model.  
To cover the entire energy range, we ran a multicanonical simulation \cite{yan, multicanonical} using the exact density of states $g^*(U)$ \cite{beale} in the sampling weight for $10^7$ MC steps per site.  
As shown in Fig. \ref{fig:isent}(a), the temperature from Eqs. (\ref{eq:mechtrans}) and (\ref{eq:gaussian}) behaved badly 
except around $U \approx 0$ or  $\beta_U \approx 0$, 
while the integral relations 
(\ref{eq:expu}), 
(\ref{eq:symu}) and 
(\ref{eq:halfu}) 
agreed with the reference 
$\beta^*(U)$.
%, which was computed by numerically differentiating  $g^*(U)$.  
This is expected, as the smallest energy change is 4 in the system, 
$\beta \, \eps \ll 1$ rarely holds, 
while $\eps$ is still $O(1)$ to justify the use of the integral relations.  
But even in the latter case, there was an $O(1)$ difference between %the logarithmic density of states from the integral 
the integral $\log g(U) = \int^U \beta_U(U')\, dU'$ 
and the exact $\log g^*(U)$ [Fig. \ref{fig:isent}(b)], 
showing that the temperature from 
Eqs. (\ref{eq:expu}), 
(\ref{eq:symu}) 
and (\ref{eq:halfu}) 
had an $O(1/N)$ systematic error without the corrections, 
as the entire potential-energy range was $O(N)$.  
The corrections (\ref{eq:expuc}), (\ref{eq:symuc}) and (\ref{eq:halfuc}), 
however, effectively removed the remaining errors
(except around the ground states, 
where the density of states was intrinsically irregular).

\subsection{\label{sec:tmatch}Temperature matching}

As an application, Eq. (\ref{eq:canon}) 
can be used to help a simplified potential-energy function 
emulate a more complex or realistic one \cite{coarsegrain, forcematching, shell}.  
In the following example, 
the simplified function is the hard-sphere potential $U_{\hs}(\vct{q})$, and the more realistic one is the LJ potential $U_{\LJ}(\vct{q})$.  
%Let us consider the following protocol.
We first run a simulation using $U_{\hs}$ as the potential energy.  
From the trajectory, 
we evaluate the effective temperature $\beta_{\LJ}$, 
by seeking the solution of 
$\la e^{-\beta_{\LJ} \Delta U_{\LJ}} \ra_{\hs} = 1$, 
with $\Delta U_{\LJ}$ being the change %of $U_{\LJ}$ 
caused by a virtual perturbation.  
We repeat the process with a modified 
%parameters of   
$U_{\hs}$
until the $\beta_{\LJ}$ matches the simulation temperature $\beta$:
% of the hard sphere system:
\begin{equation}
\la e^{-\beta_{\LJ} \, \Delta U_{\LJ}} \ra_{\hs} = 1.
\label{eq:tmatchlj}
\end{equation}
Thus, Eq. (\ref{eq:tmatchlj}) calibrates the hard-sphere system by the virtual thermometer gauged in the LJ system (cf. Appendix \ref{apd:relentropy}).

Similar to Eq. (\ref{eq:tmatchlj}), one can also match the potential energy as 
$\la U_{\LJ} \ra_{\hs} = \la U_{\LJ} \ra_{\LJ}$
or pressure as
$\la p_{\LJ} \ra_{\hs} = \la p_{\LJ} \ra_{\LJ}$.  
Another alternative is to minimize the relative entropy \cite{hansen, shell, gbproof}
\begin{equation}
S_r =
\int \log[w_{\hs}(\vct{q}) / w_{\LJ}(\vct{q})]
		w_{\hs}(\vct{q}) \, d\vct{q},
\label{eq:relentlj}
\end{equation}
which measures the difference between the distributions
$w_{\LJ} = e^{-\beta U_{\LJ}}/Z_{\LJ}$
 and 
$w_{\hs} = e^{-\beta U_{\hs}}/Z_{\hs}$.  
Although Eq. (\ref{eq:tmatchlj}) and the conditions of 
matching the potential energy and pressure can 
all be derived from minimizing $S_r$ regarding 
virtual variations \cite{hansen, shell, gbproof} of $w_{\LJ}$ (Appendix \ref{apd:relentropy}), 
they do not necessarily find the exact optimal $U_{\hs}$ that minimizes $S_r$.

The methods were tested on a fluid system of 108 particles.  
The diameter $r_a$ of the hard-sphere system 
was varied to match the LJ system under three conditions.  
The ratio of the partition functions 
$Z_{\LJ}/Z_{\hs}$, 
required for computing $S_r$, 
were obtained by Bennett's acceptance ratio method \cite{bennett} 
from independent simulations on the two systems.  
Table \ref{tab:tmatch} shows that the obtained $r_a$ from the above methods were generally close in the gaseous phase.  
It was, however, not always possible to find solutions for all methods, 
e.g., matching the potential energy failed in the second case, 
while the relative entropy was hard to compute in the third case.  
Thus, the methods can be complementary.
Like the force-matching method \cite{forcematching},
the temperature matching
requires simulation in only one (hard-sphere) system.
One may also treat the configuration temperature as a special thermal force, 
and thus add a restraint to the force-matching method.  
Also note that Eq. (\ref{eq:tmatchlj}) allows
a local perturbation with cheap local energy calculations.

\begin{table}[h]\footnotesize
\caption{
Matching of the hard-sphere and Lennard-Jones systems.
}
\centering
\begin{tabularx}{\linewidth}
{*{7}{Y}}
\hline
$r_a^\dagger$ & 
$S_r(k_B)^\ddagger$ & 
$\beta_{\LJ}^\P$ & 
$\la U_{\LJ} \ra_{\hs}^\S$ & 
$\la U_{\LJ} \ra_{\LJ}^\S$ & 
$\la p_{\LJ} \ra_{\hs}^\S$ & 
$\la p_{\LJ} \ra_{\LJ}^\S$ 
\\ 
\hline 
\multicolumn{7}{c}{
I. $T = 2.5$, $\rho = 0.3$
}\\
0.900	&15.2	&0.24	&$-1.25$	&$-1.82$	&1.620	&0.728 \\
0.952	&\textbf{7.1}	&0.62	&$-1.73$	&$-1.82$	&\textbf{0.729}	&\textbf{0.729} \\
0.970	&8.4	&0.86	&$\mathbf{-1.82}$	&$\mathbf{-1.82}$	&0.532	&0.728 \\
0.978	&9.4	&\textbf{1.00}	&$-1.84$	&$-1.82$	&0.459	&0.727 \\
1.000	&13.6	&1.52	&$-1.90$	&$-1.82$	&0.289	&0.728 \\
\hline
\multicolumn{7}{c}{
II. $T = 1.0$, $\rho = 0.05$
}\\
0.900 & 13.4 & 0.23 & $-0.21$ & $-0.47$ & 0.065 & 0.036 \\
0.981	& $\mathbf{8.6}$	& $\mathbf{1.01}$	& $-0.28$	& $-0.47$	& 0.038	& 0.036 \\
0.993	 & 8.8	& 1.27	&$-0.28$	& $-0.47$	& \textbf{0.036}	& \textbf{0.036} \\
1.000	& 9.0	& 1.40	& $-0.28$	& $-0.47$	& 0.035	& 0.036 \\
\hline
\multicolumn{7}{c}{
III. $T = 1.0$, $\rho = 0.7$
}\\
0.900	& 180	& 0.28	& $-2.33$	& $-4.89$	& 9.88	& $-0.06$ \\
0.970	& 37	& \textbf{1.00}	& $-4.44$	& $-4.89$	& 1.89	&$-0.05$ \\
0.999	&-	& 1.74	& $-4.88$	& $-4.90$	& $\mathbf{-0.04}$	& $\mathbf{-0.06}$ \\
1.000	&-	& 1.77	& $\mathbf{-4.89}$	& $\mathbf{-4.89}$	& $-0.11$	& $-0.06$ \\
\hline
\multicolumn{7}{p{\linewidth}}{
Each simulation took $10^7$ MC steps;
$r_c = 2.5$ for the LJ potential.  
The points of minimal entropy and matched quantities are shown in boldface.
}\\
\multicolumn{7}{p{\linewidth}}{
$^\dagger$ 
Diameter of hard spheres.
}\\
\multicolumn{7}{p{\linewidth}}{
$^\ddagger$ 
Defined in Eq. (\ref{eq:relentlj}).
}\\
\multicolumn{7}{p{\linewidth}}{
$^\P$ From Eq. (\ref{eq:tmatchlj}) with a single-particle perturbation and $u_{\max} = 0.02$.
}\\
\multicolumn{7}{p{\linewidth}}{
$^\S$ Figures have been divided by the number of particles.
}\\
\hline
\end{tabularx}
\vspace{-0.1in}
\label{tab:tmatch}
\end{table}

\section{Conclusions}

	To sum up, temperature   can be extracted from the distribution $p(\eps)$  of the potential-energy change $\eps$  caused by configuration perturbations, as the nontrivial root of Eqs. (\ref{eq:expu}), 
	(\ref{eq:symu}) or (\ref{eq:halfu}).  
The formulas can be understood as virtual thermometers gauged in a corresponding canonical ensemble.
When used in the constant potential-energy ensemble, the formulas have an $O(1/N)$ error, but can be corrected systematically.

The approach can be extended to the mean force of a multidimensional distribution $\rho(\vec{X})$ by an ensemble-adjusted perturbation.  The adjusted perturbation is equivalent to a short trajectory of a reversible dynamics, making the mean force formulas (\ref{eq:expx}) and (\ref{eq:expxmd}) readily usable in simulations and experiments \cite{experiments}.
The computer code of the examples can be found in Ref. \cite{software}.

\section*{Acknowledgements}
	I thank Dr. M. W. Deem, Dr. B. M. Pettitt, and the referees for helpful discussions and comments.  I am especially indebted to the second referee for various insightful suggestions on, among others, the non-equilibrium extensions, 
	connections with coarse-graining methods, and verifications on experiments and exact models.

\appendix
\section{\label{apd:ctemp} 
Configuration temperature}

%We derive Eq. (\ref{eq:ctemp}) in the constant potential-energy ensemble (defined in the Sec. \ref{sec:uensemble}).  
Following Refs. \cite{rugh, butler, jepps},
we show that the temperature $\beta_U = \frac{d}{dU}g(U)$ in the constant potential-energy ensemble (cf. Sec. \ref{sec:uensemble})
%(with $g(U) = \int \delta[U - U(\vct{q})] \, d\vct{q}$)
satisfies
\begin{align}
\beta_U
= \frac{\la \nabla \cdot \vct{B}\ra_U}
{\la \vct{B} \cdot \nabla U \ra_U}
-
\frac{d \log\la \vct{B} \cdot \nabla U\ra_U}
{dU},
\label{eq:ctempu}
\end{align}
where $\vct{B} = \vct{B}(\vct{q})$ is a vector field 
that satisfies $\vct{B}\cdot \nabla U > 0$ \cite{jepps}.  The second term on the right is $O(1/N)$. We first define \cite{jepps}
\[
G(U)
= 
\int (\vct{B} \cdot \nabla U) \,
\delta[ U(\vct q) - U] \, d\vct q
=
\la \vct B \cdot \nabla U \ra_U \, g(U).
\]
Then
\begin{align}
\frac{dG}{dU}
= \left(
\frac{d \la \vct{B} \cdot \nabla U \ra_U}{dU}
+ \la \vct{B} \cdot \nabla U \ra_U \beta_U 
\right) g(U).
\label{eq:dGdUa}
\end{align}
On the other hand, integration by parts yields
\begin{align}
\frac{dG}{dU}
&= -\int
  \vct{B} \cdot \nabla \delta[U(\vct{q}) - U] d \vct{q} \notag \\
&= \int (\nabla \cdot \vct{B}) \delta[U(\vct{q}) - U] d \vct{q}
  = \la \nabla \cdot \vct{B} \ra_U g(U).
\label{eq:dGdUb}
\end{align}
Equating Eqs. (\ref{eq:dGdUa}) and (\ref{eq:dGdUb}) gives (\ref{eq:ctempu}).
Eq. (\ref{eq:ctemp}) represents two special cases.  
With $\vct B = \nabla U/(\nabla U \cdot \nabla U)$, 
$\vct B \cdot \nabla U = 1$, and
\begin{align}
\beta_U 
%&= \left\la \nabla \cdot \left( 
%		\frac{\nabla U}{\nabla U \cdot \nabla U} %\right) \right\ra_U 
%		\notag \\
&= \left\la \frac{\nabla U} {\nabla U \cdot \nabla U} \right\ra_U
	-\left\la 
	\frac{2 \nabla U \cdot \nabla \nabla U \cdot \nabla U }
	{(\nabla U \cdot \nabla U)^2} 
	\right\ra_U,
\label{eq:ctempu1}
\end{align}
where the second term is $O(1/N)$.  
With $\vct{B} = \nabla U$, we get
\begin{equation}
\beta_U 
= \frac{\la \nabla^2 U \ra_U}
	{\la \nabla U \cdot \nabla U \ra_U}
- \frac{d \log \la \nabla U \cdot \nabla U\ra_U} {d U}.
\label{eq:ctempu2}
\end{equation}
%
%Although both Eqs. (\ref{eq:ctempu1}) and (\ref{eq:ctempu2}) require the second derivatives of $U(\vct{q})$, the latter requires the Laplacian $\nabla^2 U$ only, 
%while the former requires 
%$\nabla U \cdot \nabla \nabla U \cdot \nabla U$
%as well.

%Without the restriction of a fixed $U$, 
We can show Eq. (\ref{eq:ctemp}) 
from the equation of motion \cite{darve2008, hansen}:
\begin{align*}
\langle \vct{B} \cdot \nabla U \rangle
&= \frac{1}{\tau}
\int_0^\tau (\vct{B} \cdot \nabla U) \, dt
= -\frac{1}{\tau}
\int_0^\tau (\vct{B} \cdot \dot{\mathbf p}) \, dt \\
&= \frac{1}{\tau}
\int_0^\tau (\vct{p} \cdot \dot{\vct{B}}) \, dt 
= \frac{1}{\tau}
\int_0^\tau ( \vct{p} \cdot \nabla \vct{B} \cdot \dot{\vct{q}}) \, dt \\
&= \la \vct{p} \cdot \nabla \vct{B} \cdot \vct{p}/m \rangle
\approx
k_B T \la \nabla \cdot \vct{B} \ra,
\end{align*}
where $\tau$  is the length of a long trajectory.  
We have assumed a thermostat \cite{hoover, thermostat2}
that makes momenta $\vct{p}$ independent:  
$\la p_i p_j/m\ra = k_B T \delta_{ij}$, 
and uncorrelated with  
$\vct{B}$ and $\nabla \vct{B}$.  
This derivation has the advantage of replacing the strong assumption of global ergodicity in the ensemble theory by a weaker condition of sufficiently randomized momenta (which requires only local equilibration).

\section{\label{apd:fluc}
Fluctuation theorems}

Eq. (\ref{eq:canon}) can be generalized by several fluctuation theorems 
\cite{maes, bkfluc, fluc, jarzynski, kurchan, seifert2005}.  
For any  $\beta'$, we have
\[
\la \exp[(\beta' - \beta) \, \eps] \ra_\beta 
= \la \exp(-\beta' \eps^*) \ra_\beta,
\]
where $\eps$ and  $\eps^*$ are the energy changes caused by a pair of uniform perturbations $+\vct u$ and $-\vct u$, respectively \cite{maes}, because
\begin{align*}
\int e^{
	(\beta' - \beta)\, [U(\vct{q} + \vct{u}) - U(\vct{q})] } \,
	e^{-\beta U(\vct{q})}/Z(\beta) \, d\vct{q} \\
=
\int e^{
	-\beta' \, [U(\vct{q}'-\vct{u}) - U(\vct{q}')] } \,
	e^{-\beta U(\vct{q}')}/Z(\beta) \, d\vct{q}',
\end{align*}
where $\vct{q}' = \vct{q} + \vct{u}$.  
For a symmetrically randomized perturbation, 
%in which $+\vct{q}$ and $-\vct{q}$ are equally likely, 
the distributions of $\eps$ and $\eps^*$  are identical, and
\begin{equation}
\la \exp[(\beta' - \beta) \, \eps] \ra_\beta
= \la \exp(-\beta' \eps) \ra_\beta.
\label{eq:canonbp}
\end{equation}
The $\eps$ distribution can be found from the exponential average 
$g(i\omega) \equiv \la \exp(-i \omega \eps) \ra_\beta$ \cite{vankampen} as
$p(\eps) = 
\frac{1}{2\pi}
\int_{-\infty}^{\infty} g(i \omega) \, e^{i\omega \eps} \, d\omega$.
By taking the Fourier transform of Eq. (\ref{eq:canonbp}), we get
%Since $G(\omega) = G(-i\beta - \omega)$ [by Eq. (\ref{eq:canonbp})], we get
\begin{equation}
p(\eps) \, \exp( -\beta\eps) = p(-\eps),
\label{eq:canondb}
\end{equation}
assuming that $g(i \omega)$ has no singularity in the strip 
$0 < \mathrm{Re} \, (i \omega) < \beta$
of the complex plane.
%Eqs. (\ref{eq:canonbp}) and (\ref{eq:canondb}) are verified on a harmonic oscillator in Appendix \ref{apd:ho}.  
Note that while Eqs. (\ref{eq:canonbp}) and (\ref{eq:canondb}) 
require a symmetric perturbation, Eq. (\ref{eq:canon}) does not.

Eq. (\ref{eq:canon}) has a few other generalizations.  
For a Hamiltonian $H_\lambda(\vct{q})$ parameterized by $\lambda$, 
we can treat a circular switch of 
$\lambda: 0 \rightarrow 1 \rightarrow 0$
starting from the canonical equilibrium state at $\lambda = 0$
as an elaborate perturbation.  
%Since the switch produces no free energy difference ($\Delta F = 0$), 
Then the Jarzynski equality states $\la \exp(-\beta W) \ra = 1$  for the non-equilibrium work 
$W = \int_0^\tau (\partial H_\lambda/\partial \lambda) \, \dot \lambda \, dt$  over a period $\tau$ \cite{jarzynski}.  
Similarly, we have $\la \exp(-\beta \tilde W) \ra = 1$
for the work
$\tilde{W} = \int_0^\tau \vct{f}(\vct{q}, t) \cdot \dot{ \vct{q}} \, dt$
derived from a time-dependent driving force $\vct{f}(\vct{q}, t)$ 
(excluding the component from the conservative potential)
\cite{bkfluc, seifert2008}.
These relations can also be used to extract the equilibrium temperature $\beta$ of the initial equilibrium state.

\section{\label{apd:ho}
Harmonic oscillator}

We verify a few formulas for the canonical ensemble on a $D$-dimensional harmonic oscillator with the potential energy $U(\vct{q}) = \frac{1}{2}k \vct{q}^2$.
For a Gaussian perturbation  
$p_\vct{u}(\vct{u}) \propto
 \exp(-\vct{u}^2/2\sigma^2)$ 
applied to a fixed $\vct{q}$, 
the average
%we define $g_\vct{q}(b) \equiv %\overline{\exp(-b \, \eps)}$, and
\begin{align*}
g_\vct{q}(b) 
&\equiv \overline{\exp(-b\, \eps)}
=
\int e^{-b[U(\vct{q}+\vct{u}) - U(\vct{q})]}
p_\vct{u} (\vct{u}) \, d\vct{u} \\
&= (1+\alpha b)^{-D/2}
\exp[\alpha b^2 U /(1+\alpha b)],
\end{align*}
where $\alpha = k \sigma^2$.  
Averaging $g_\vct{q}(b)$ over $\vct{q}$ yields
\begin{align}
g(b) 
&\equiv\la g_{\vct{q}}(b) \ra_\beta
= \int g_{\vct{q}}(b) \, e^{-\beta U(\vct{q})}/Z(\beta) \, d\vct{q}
\notag\\
&=[1 + \alpha b(1-b/\beta)]^{-D/2}.
\label{eq:hogenf}
\end{align}
Eq. (\ref{eq:hogenf}) 
satisfies Eqs. (\ref{eq:canon}) and (\ref{eq:canonbp}), i.e., $g(\beta') = g(\beta - \beta')$.

Since $g(b)$ gives the moment generating function \cite{vankampen}, the $\eps$ distribution $p(\eps)=\frac{1}{2\pi} \int_{-\infty}^{+\infty} g(i\omega) e^{i\omega \eps} \, d\omega$ is
\begin{equation}
p(\eps) = 
\dfrac
{[|\eps|/(2\gamma)]^{(D-1)/2}}{(\alpha/\beta)^{D/2}}
\dfrac
{K_{(D-1)/2}(\gamma |\eps|)}
{\sqrt\pi \Gamma(D/2)}
e^{\beta \eps/2},
\label{eq:hope}
\end{equation}
where $\gamma = \sqrt{(\beta/2)^2 +(\beta/\alpha)}$,
$K_n(x)$ is the modified Bessel function of the second kind, 
and $\Gamma(x)$ is the gamma function.  
Eq. (\ref{eq:hope}) satisfies Eq. (\ref{eq:canondb}), 
and shows that $p(\eps)$ at $\eps = 0$ depends critically on the dimension $D$: 
it diverges if $D < 2$; it has a finite cusp if  $D = 2$; and it is differentiable if $D > 2$.
In the limit of $D \gg 1$  and  $\alpha \beta \ll 1$, the distribution is Gaussian:
\begin{equation}
p(\eps) \approx \big(1/\sqrt{2\pi \sigma^2}\big)
\exp\big[-(\eps - \alpha D/2)^2/(2\sigma^2)\big],
\label{eq:hogaussian}
\end{equation}
where $\sigma^2 =(\alpha/\beta + \alpha^2/2)D
\approx (\alpha/\beta) D$.  
%The limiting distribution is Gaussian because $\eps$ is the sum of independent one-dimensional perturbation energy. 

\section{\label{apd:series}
Series expansion}

We derive the corrections for 
Eqs. (\ref{eq:expu}), (\ref{eq:symu}) and (\ref{eq:halfu}) 
from
\begin{align}
p_U(-\eps)
&= \dfrac{g(U-\eps)}{g(U)} p_{U-\eps}(+\eps)
\notag\\
&=
\exp\left[ \sum_{l=1}^\infty
\dfrac{b_l(-\eps)^l} {l!}\right]
\sum_{m = 0}^{+\infty}
\dfrac{(-\eps)^m}{m!}
\dfrac{d^m p_U(\eps)} {dU^m},
\label{eq:series}
\end{align}
where $b_l = \frac{d^l}{d U^l} \log g(U)$.  
Eq. (\ref{eq:expu}) seeks the root $\hat\beta$ of
\[
0 = \int_{-\infty}^{+\infty}
\left[ p_U(-\eps) \, e^{\hat \beta \, \eps} - p_U(\eps) \right] \,
	\eps^k \,d \eps.
\]
Using Eq. (\ref{eq:series}) for $p_U(-\eps)$ yields:
\begin{equation}
\langle \eps^k \rangle_U 
=
\sum_{m = 0}^{\infty}
\tfrac{ (-)^m } {m !}
\tfrac{ \bar d^m} {\bar d U^m}
\left\langle
	\exp\left[ 
		\hat \beta \eps +  
		\textstyle\sum_l\tfrac{b_l (- \eps)^l}{l!}  
	\right]
	\eps^{k+m}
\right\rangle_U, 
\label{eq:expuav}
\end{equation}
where $\bar d^m/\bar d U^m$ denotes a differentiation 
that applies only to the moments of $\eps$, 
but not to the $\hat \beta$ or $b_l$.

Eq. (\ref{eq:expuav}) can be solved by successive approximations:
$\hat \beta = \hat \beta^{(1)} + \hat \beta^{(2)} + \dots$
with  $\hat \beta^{(n)} \sim O(1/{N^{n-1}})$.
To the first order, 
we use $\hat\beta = \hat\beta^{(1)}$ 
in Eq. (\ref{eq:expuav}), and
\begin{equation*}
\langle \eps^k \rangle_U 
\approx 
\langle \exp(\hat \beta^{(1)}\eps - b_1 \eps)  \eps^k \rangle_U,
\end{equation*}
which yields $\hat \beta^{(1)} = b_1$.  
%To the second order, 
Next, we set $\hat \beta = \hat \beta^{(1)} + \beta^{(2)}$, and
\begin{align*}
\left\langle
	\eps^k 
\right\rangle_U 
&\approx 
\left\langle \exp\left(
	\hat \beta^{(2)}\eps + \tfrac{1}{2} b_2 \eps
	\right)  \eps^k 
\right\rangle_U  
-\tfrac{d}{dU} \left\langle \eps^{k+1} \right\rangle_U,
\end{align*}
which yields 
$\hat\beta^{(2)} 
= \left[
	-\frac{1}{2}\, b_2 \la \eps^{k+2} \ra_U
+ \frac{d }{dU} \la \eps^{k+1} \ra_U 
	\right]/ {\la \eps^{k+1} \ra_U}$.
We reach Eq. (\ref{eq:expuc}) by $d\beta_U/dU \approx d\hat \beta/dU$.

Eq. (\ref{eq:expuc}) is exact for any $N$ in the limit of small $\eps$.
For an even $k$ in this limit, it becomes
(cf. Sec. \ref{sec:mechtrans})
\begin{equation}
\beta_U = 
	\dfrac { 2\la \eps^{k+1} \ra_U }
				{ \la \eps^{k+2}\ra_U  }
	- \dfrac{d \log \la \eps^{k+2} \ra_U}{dU},
\label{eq:expucsmalle}
\end{equation}
with
$\hat\beta = 2\la \eps^{k+1} \ra_U/\la \eps^{k+2} \ra_U$.
To show this, we use
\begin{align*}
\,\la \eps^{k+2} \rangle_U
&=
\textstyle\sum_{\{i\}}
\la \partial_{i_1} U \cdots \partial_{i_{k+2}} U \ra_U \,
	\overline{u_{i_1}\cdots u_{i_{k+2}}}, \\
2 \langle \eps^{k+1} \ra_U 
&=
(k+1)
\textstyle\sum_{\{i\}}
\la \partial_{i_1} U \dots \partial_{i_{k}} U 
	\partial_{i_{k+1}i_{k+2}}^2 U \ra_U \,
	\overline{u_{i_1} \dots u_{i_{k+2}}}.
\end{align*}
Partial integration yields 
$\frac{d}{dU}
[\la \eps^{k+2} \ra_U \, g(U)] =
2 \la \eps^{k+1} \ra_U \, g(U)$, 
which is Eq. (\ref{eq:expucsmalle}).  
For an odd  $k$, the  $\beta_U$  value is identical to that from the $k-1$ case in the small $\eps$ limit.

The correction for Eq. (\ref{eq:symu}) can be constructed from
\[
0 = \int_{-\infty}^{\infty}
\sgn(\eps) |\eps|^k
\min\left\{1, e^{-\hat\beta \,\eps } \right\}
\left[
	p_U(-\eps)\,e^{\hat\beta \,\eps} - p_U(\eps)
\right]\,
d\eps.
\]
Using Eq. (\ref{eq:series}) for $p_U(\eps)$ yields
\begin{equation}
\beta_U \approx
\hat \beta
-	\dfrac{m_{k+2}}{2 \, m_{k+1}} 
	\dfrac{d |\hat \beta|}{dU} 
-	\dfrac{d \log m_{k+1}}{dU},
\label{eq:symuc}
\end{equation}
where $m_l \equiv \int_{-\infty}^{+\infty} 
|\eps|^l \min\{1, \exp(-\hat{\beta} \, \eps)\} p(\eps) \, d\eps$.
%and $\hat \beta$ is the value determined from Eq. (\ref{eq:symu}).  
To correct the $\hat \beta$ value determined from Eq. (\ref{eq:halfu}), 
we use
\[
0 = \int_{-\infty}^{\infty}
\sgn(\eps) |\eps|^k
e^{-\hat \beta \eps/2}
\left[
	p_U(-\eps)\,e^{\hat\beta \eps} - p_U(\eps)
\right]\,
d\eps,
\]
which yields
\begin{equation}
\beta_U \approx
\hat \beta
- \dfrac{d}{dU} \log 
	\left\la |\eps|^{k+1} 
		\exp\left(
			-\dfrac{1}{2} \hat\beta \eps
		\right) 
	\right\ra_U,
\label{eq:halfuc}
\end{equation}
Eq. (\ref{eq:halfuc}) lacks the $d\hat\beta/dU$ term in 
Eqs. (\ref{eq:expuc}) and (\ref{eq:symuc}),
as it
has been included in differentiating the 
$\hat\beta$ in the %factor 
$\exp(-\frac{1}{2}\hat\beta\,\eps)$.

\section{\label{apd:meanforce}Distribution mean force}

%	For the distribution $\rho(X)$ defined by Eq. (\ref{eq:distr1d}), we have
%
%
%\begin{align*}
%\frac{d\log \rho}{d X}
%= \frac{\la \nabla \cdot \vct{B} + \vct{B} \cdot \nabla \log w\ra_X}
%{\la \vct{B} \cdot \nabla X \ra_X}
%-
%\frac{d \log\la \vct{B} \cdot \nabla X\ra_X}
%{dX}.
%\end{align*}
%
%
%where $\vct{B}$ is a vector field that satisfies $\vct{B}\cdot\nabla X > 0$. 
%and $\la\dots\ra_X$ denotes an average under a fixed $X$ weighted by $w(\vct{q})$.
%The proof is similar to that of Eq. (\ref{eq:ctempu}).

For the $K$-dimensional  
distribution \cite{rugh, meanforce, bluemoon, darve2008, zhang} $\rho(\vec X)$
% $\vec{X} = \{X_\alpha\}$ 
%($\alpha = 1, \dots, K$),
defined by Eq. (\ref{eq:distrmd}),  
we have
\begin{equation}
  f_\alpha
  \equiv \frac{\partial \log \rho}{\partial X_\alpha} 
  = \sum_\gamma (\mat{W}^{-1})_{\alpha\gamma}
  \left(F_\gamma - 
   \sum_\theta \dfrac{\partial W_{\gamma \theta}}{\partial X_\theta} 
   \right).
\label{eq:mfmd}
\end{equation}
where $\mat W$ is the $K$  by $K$ matrix with
$W_{\alpha \gamma}(\vec{X})
\equiv
\la \vct{B}_\alpha \cdot \nabla X_\gamma \ra_{\vec X}$
(the $\vct{B}_\alpha$
are vector fields such that 
the matrix has an inverse
${\mat W}^{-1}$), and
$F_\alpha \equiv 
\la \nabla \cdot \vct{B}_\alpha
 + \vct{B}_\alpha \cdot \nabla \log w \ra_{\vec X}.
$
Particularly, if $K = 1$,
\begin{align*}
\frac{d\log \rho}{d X}
= \frac{\la \nabla \cdot \vct{B} + \vct{B} \cdot \nabla \log w\ra_X}
{\la \vct{B} \cdot \nabla X \ra_X}
-
\frac{d \log\la \vct{B} \cdot \nabla X\ra_X}
{dX}.
\end{align*}
which is similar to Eq. (\ref{eq:ctempu}).
To show Eq. (\ref{eq:mfmd}), we observe
\[
W_{\alpha \gamma} \, \rho
=
\int
(\vct{B}_\alpha \cdot \nabla X_\gamma) \,
\textstyle\prod_\gamma
\delta[X_\gamma(\vct{q}) - X_\gamma]
w(\vct{q})\,d\vct{q}.
\]
Then partial integration gives
\begin{align*}
F_\alpha \rho
&=
\int \nabla \cdot[w(\vct{q}) \, \vct{B}_\alpha] \,
\textstyle\prod_{\gamma}
	\delta[X_\gamma(\vct{q}) - X_\gamma] \,
	d \vct{q} \\
&= \textstyle\sum_\gamma 
\partial{(W_{\alpha\gamma}\,\rho)}/{\partial X_\gamma}
=
\textstyle\sum_\gamma 
\left( W_{\alpha \gamma} f_\gamma + 
\partial{W_{\alpha\gamma}}/
{\partial X_\gamma} \right)\, \rho.
\end{align*}
Multiplying $\mat{W}^{-1}$ to both sides yields Eq. (\ref{eq:mfmd}).

%We discuss two special cases of Eq. (\ref{eq:mfmd}).  
%In the first case,  
If $\vct{B}_\alpha$ is constructed by the Gram-Schmidt orthonormalization such that   
$W_{\alpha\gamma} = 
\la \vct{B}_\alpha \cdot \nabla X_\gamma \ra_{\vec{X}}
= \delta_{\alpha \gamma}$
\cite{rugh, bluemoon}, then
\[
f_\alpha
=
F_\alpha = 
\la 
\nabla \cdot \vct{B}_\alpha
+\vct{B}_\alpha \cdot \nabla \log w
\ra_{\vec X}.
\]	
The resulting $\nabla \cdot \vct{B}_\alpha$ can be quite complex \cite{henin},
e.g., for $K = 2$,
\[
\vct{B}_\alpha
=
\frac{
\textstyle\sum_{\mu=1,2} 
\nabla X_\mu \cdot \nabla X_\mu 
-\nabla X_\mu \otimes \nabla X_\mu 
}{
(\nabla X_1 \cdot \nabla X_1)
(\nabla X_2 \cdot \nabla X_2)
-(\nabla X_1 \cdot \nabla X_2)^2
} \, \nabla X_\alpha.
\]
We can alternatively set $\vct{B}_\alpha = \nabla X_\alpha$, and
\begin{equation}
\begin{split}
W_{\alpha \gamma}(\vec{X})
&= \la \nabla X_\alpha \cdot \nabla X_\gamma \ra_{\vec X}, \\
F_\alpha 
&= \la \nabla^2 X_\alpha
+ \nabla X_\alpha \cdot 
\nabla \log w \ra_{\vec X},
\label{eq:mfmd2}
\end{split}
\end{equation}
which limits the second derivatives to the Laplacians,
but requires the corrections
$\frac{\partial}{\partial X_\theta} W_{\gamma \theta}$
from numerical differentiation.
%The advantage is that the only second derivatives required are the Laplacians $\nabla^2 X_\alpha$.  
%The corrections 
%$\frac{\partial}{\partial X_\theta} W_{\gamma \theta}$ 
%can be computed by numerical differentiation.

\section{\label{apd:relentropy}Relative entropy}

%	The relative entropy $S_r$  of two distributions 
%	$w_A(\vct{q})$ and $w_B(\vct{q})$ is defined as
%\begin{equation}
%S_r = \left\langle \log \left( 
%	\dfrac{w_B}{w_A} \right) \right\rangle_B
%	=-\int \log\left[\dfrac{w_A(\vct{q})}
%	{w_B(\vct{q})}\right] w_B(\vct{q}) \, d\vct{q},
%\label{eq:relentropy}
%\end{equation}
%whose minimum is located at $w_A = w_B$ 
%[use $-\log x \ge 1 - x$ 
%with $x = w_A/w_B$ in Eq. (\ref{eq:relentropy})\cite{hansen, shell, gbproof}].

We show below that the temperature matching condition Eq. (\ref{eq:tmatchlj})
locally minimizes the relative entropy \cite{shell}
\[
S_r^*
	=-\int \log\left[\frac{w_A(\vct{q}')}{w_B(\vct{q})}
	\left|\frac{\partial \vct{q}'}{\partial \vct{q}}
	\right|
	\right] \, w_B(\vct{q}) \, d\vct{q},
\]
of two distributions 
$w_A(\vct{q})$ and $w_B(\vct{q})$
 ($A\rightarrow \LJ, B\rightarrow \hs$) along the force.
If $\vct{q}' = \vct{q}$, the minimum is located at $w_A = w_B$ 
[use $-\log x \ge 1 - x$ 
with $x = w_A/w_B$ \cite{hansen, shell, gbproof}].
If, however, $S_r^*$ is minimized along an infinitesimal coordinate transformation 
$\vct{q}' = \vct{q} + \delta\lambda \vct{B}(\vct{q})$,
then $\partial S_r^*/\partial(\delta \lambda) = 0$, or
$\la \nabla \cdot \vct{B} + \vct{B} \cdot \nabla \log w_A \ra_B = 0$.
In the canonical ensemble,  
\begin{equation}
	\beta_A = 
	\left\la \nabla \cdot \vct{B} \right\ra_B / 
	\left\la \vct{B} \cdot \nabla U_A \right\ra_B,
\label{eq:beteff}
\end{equation}
for $w_A(\vct{q}') \propto \exp[-\beta_A U_A(\vct{q}')]$.
Particularly, with $\vct B = \nabla U_A$, 
we have $\beta_A = \la \nabla^2 U_A \ra_B / \la \nabla U_A \cdot \nabla U_A \ra_B$, which is Eq. (\ref{eq:tmatchlj}) in the small-perturbation limit.
%cf. the transition from Eq. (\ref{eq:mechtrans}) to Eq. (\ref{eq:canon}).  
Generally, Eq. (\ref{eq:beteff}) defines a distinct effective temperature $\beta_A$ for each vector field $\vct{B}$,
reflecting the fact that one can define
different effective temperatures \cite{casas}
in a non-equilibrium state 
($w_A$: equilibrium, $w_B$: non-equilibrium).
The conditions for matching the potential energy and pressure can be similarly obtained by varying  $w_A$ with respect to the temperature 
and volume, respectively.

Normally, we set $w_A$ as the model distribution, and optimize its parameters to match the reference $w_B$
\cite{shell};
thus, the averages should be performed in the reference $B$ system.  This is, however, inconvenient, for the example in Sec. \ref{sec:tmatch}, since $\log w_{\hs}$, 
when averaged over configurations produced 
by the reference LJ potential, 
can be infinite, 
due to the hard-sphere potential $U_B$.  
We therefore set $A \rightarrow \LJ$ and
 $B \rightarrow \hs$.

\section{\label{apd:langevin}Driven Langevin system}

We use a driven system in constant contact with a heat bath as a model to study the temperature in a non-equilibrium steady state.  We will show below that, to correctly extract the presumably constant heat bath temperature, Eq. (\ref{eq:canon}) should be applied to a nonuniform perturbation.

Consider the overdamped Langevin equation 
\cite{kurchan, seifert2005}:
$
d\vct{q}/dt = \vct{F}/\eta + \sqrt{2/(\beta\eta)} \xi$,
where the total force $\vct{F}$ is the conservative component $-\nabla U$ plus a driving $\vct{f}$, $\eta$ is the viscosity, and $\xi$ is a Gaussian white noise satisfying 
$\la \xi_i(t) \xi_j(t')\ra = \delta_{ij}
\delta(t - t')$.  
The steady-state distribution $w$ satisfies %the Fokker-Planck equation
\begin{equation}
\nabla \log w = \beta \vct{F} - \beta \eta \vct{j}/ w,
\label{eq:fp}
\end{equation}
where  $\vct j$ is the constant current, and the ratio $\vct{v} = \vct{j}/w$ is the average local velocity \cite{seifert2005}.  
The equilibrium case $w\propto \exp(-\beta U)$ is recovered if $\vct{f} = \vct{j} = \vct{0}$.

Consider a perturbation $\vct q \rightarrow \vct q'$ 
generated by a flow from the vector field $\vct{u}(\vct{q})$,
$d\vct{q}/d\tau = \vct{u}[\vct{q}(\tau)]$,
over a period of ``virtual time'' $\tau_m$, such that $\vct q(\tau = 0) = \vct{q}$  and  
$\vct q(\tau = \tau_m) = \vct{q}'$.
%Note that the virtual time  $\tau$ is unrelated to the real time t in Eq. (\ref{eq:lang}).
We then have, by Eq. (\ref{eq:fp}),
\begin{equation}
\left\la e^{-\beta \eps_{\mathrm{eff}}
-\beta q_{\mathrm{hk}}} \, J \right\ra 
= 
\int \frac{w(\vct{q}')} {w(\vct{q})}
\left| \frac{\partial \vct{q}'}{\partial \vct{q}} \right|
w(\vct{q}) \, d\vct{q} = 1,
\label{eq:expnej}
\end{equation}
where $\eps_{\mathrm{eff}} \equiv
 - \int_0^{\tau_m} \vct{F}[\vct{q}(\tau)] \cdot 
 \vct{u}[\vct{q}(\tau)] \, d\tau$ 
is the change of the local effective potential
%$\Delta U_{\mathrm{eff}}$
\cite{reimann} or the absorbed heat,
$q_{\mathrm{hk}} \equiv \int_0^{\tau_m}
\eta \vct{u}[\vct{q}(\tau)] \cdot \vct{v}[\vct{q}(\tau)] \, d\tau$
is the housekeeping heat 
\cite{kurchan, seifert2005, oono}, 
and $J \equiv \exp\left(
\int_0^{\tau_m} \nabla \cdot \vct{u}[\vct{q}(\tau)] \, d\tau
\right)
=
|\partial \vct{q}'/\partial \vct{q}|$ \cite{seifert2008}.

Eq. (\ref{eq:expnej}) can be made to locally resemble the equilibrium version [Eq. (\ref{eq:canon})] as
%\begin{equation}
$\left\la 
	\exp(-\beta \eps_{\mathrm{eff}}) 
\right\ra = 1$
%\label{eq:expne}
%\end{equation}
%
by a constraint $\exp(-\beta q_{\hk}) J = 1$.
The constraint can be satisfied by the vector field  
$\vct{u} = \vct{u}_0\sigma(\vct{q})$, 
where $\vct{u}_0$ is a constant 
and $\sigma(\vct{q})$ satisfies 
$\nabla \log \sigma(\vct{q})
= \beta \eta \vct{v}(\vct{q})$.
%(such that  $\nabla \cdot \vct{u} = \beta \eta \vct{u} \cdot \vct{v}$).  
The $\sigma(\mathbf{q})$ induces a nonuniform perturbation in the steady state with nonzero current.
%Further, since $d\vct q = \vct u_0 (\sigma d\tau)$, the $\sigma(\vct q)$ serves as a natural metric of the perturbation.
%
In this way, %Eq. (\ref{eq:expne}) allows 
we get
an energy-based reading of the temperature 
in the non-equilibrium steady state, 
just as in the equilibrium state, 
such that the same  $\beta$ value can be obtained for perturbations of different sizes and directions.
Similarly, if the perturbation is a short trajectory that follows the Langevin equation, or one that satisfies 
$\exp(-\beta \eps_{\mathrm{eff}})\, J = 1$,
then 
$\langle \exp( -\beta q_{\mathrm{hk}} ) \rangle = 1$ \cite{speck2005, seifert2008}
yields the same $\beta$.

%

%\end{equation}
%

\end{document}